# A Data Annotation Architecture for Semantic Applications in Virtualized Wireless Sensor Networks

Imran Khan*, Rifat Jafrin†, Fatima Zahra Errounda†, Roch Glitho†, Noël Crespi*,
Monique Morrow‡ and Paul Polakos‡
*Institut Minés-Télécom, Télécom SudParis, 91011 Evry Cedex, France
Email: imran@ieee.org, noel.crespi@it-sudparis.eu
†Dept. CIISE, Concordia University, H3G 2W1, Montreal, Canada
Email: {r_jafri, f_errou} @encs.concordia.ca, glitho@ciise.concordia.ca
‡CISCO Systems, Inc.
Email: {mmorrow, ppolakos} @cisco.com

*Abstract*—Wireless Sensor Networks (WSNs) have become very popular and are being used in many application domains (e.g. smart cities, security, gaming and agriculture). Virtualized WSNs allow the same WSN to be shared by multiple applications. Semantic applications are situation-aware and can potentially play a critical role in virtualized WSNs. However, provisioning them in such settings remains a challenge. The key reason is that semantic applications' provisioning mandates data annotation. Unfortunately it is no easy task to annotate data collected in virtualized WSNs. This paper proposes a data annotation architecture for semantic applications in virtualized heterogeneous WSNs. The architecture uses overlays as the cornerstone, and we have built a prototype in the cloud environment using Google App Engine. The early performance measurements are also presented.

*Keywords*—*Wireless Sensor Networks; Semantic Web; Domain Ontologies; WSN Virtualization; Data Annotation; Overlays*

## I. INTRODUCTION

Wireless Sensor Networks (WSNs) [1] consist of small-scale devices that allow applications to observe various physical phenomenon and then react to the reported events. However, WSN deployments are usually tailored for predefined applications with no possibility for new applications to use them concurrently. To address this, WSN virtualization that uses the concept of concurrent application tasks running on a sensor node and combines such nodes together to work for multiple applications simultaneously has gained considerable attention [2]–[4]. We have recently proposed an early architecture as a solution for WSN virtualization [5].

Typically, virtualized WSNs provide sensor data in raw format. However, classical WSN applications cannot interpret the raw sensor data and understand its context. This makes it almost impossible for their users to get the high-level details of the events and infer additional knowledge to gain situational awareness. For example, a fire monitoring application can only receive a simple fire notification without additional details for its end-user to understand the meanings and context of the fire event, e.g. event status and its location.

Semantic applications, on the other hand, allow their users to make queries such as *what is the current status of the fire?* and *what is the current location of the fire?* to get results like *initial fire* and *in a public library* respectively. Virtualized WSNs typically monitor several real-time events at the same time for different applications. Hence, some end-users of these applications may wish to know the context of specific events. This brings us to the need for a mechanism that annotates the sensor data in a virtualized WSN. Annotating sensor data in virtualized WSNs is quite challenging; since resources are scarce, virtual sensors are created on-demand and may have unpredictable lifetime.

In order to provision semantic applications we need to send additional metadata along with raw sensor data. For example, the raw sensor data for a fire monitoring application can be annotated with concepts such as *observed property* and *location*, which are temperature and, longitude and latitude, respectively, in this case. Semantic annotation has been a popular approach for this purpose. It is defined as a metadata generation and usage schema that can be used to provide new methods, as well as to extend existing ones, to access new information [6]. However, the semantic annotation process requires domain concepts and the relationships that exist between them in order to annotate data. An ontology is used to formally represent a domain, its concepts and the relationships that exist [7]. Within sensor domain, there are several efforts to develop ontologies, e.g., the Semantic Sensor Network (SSN) Ontology developed by the W3C Semantic Sensor Network Incubator Group [8] and SensorML from the Open Geospatial Consortium (OGC) [9]. SSN ontology is more general purpose because it is application domain independent and provides concepts about sensors and their observations.

This paper proposes a data annotation architecture for semantic applications in virtualized WSN environments. We extend our previous WSN virtualization architecture [5] to cater for data annotation. We develop a base ontology by extending the SSN ontology. We also develop a domain ontology for the semantic application we have prototyped. The fire monitoring semantic application receives annotated data and uses the fire domain ontology, along with a reasoner, to infer knowledge. An end-user can query over the annotated data to get the real-time information of the fire event, such as its status and location. The application is developed and deployed in the cloud using Google App Engine (GAE) and works in a heterogeneous virtualized WSN environment.

The rest of the paper is organized as follows: A motivating scenario is presented in Section II, along with a set of requirements. We discuss our proposed architecture in Section III, followed by our procedures and illustrative scenario in Section IV. The prototype implementation and results are discussed in Section V and an overview of the related work in Section VI. Section VII presents lessons learned along with the future work and Section VIII concludes the paper.

## II. MOTIVATING SCENARIO AND REQUIREMENTS

In this section, we first present a motivating scenario and then derive a set of requirements from it.

### A. Motivating Scenario

We extend the motivating scenario presented in [5] for a semantic application that monitors fire events in real time.

Consider a city near an area where brush fires are common and where some houses already have their own sensors to detect fire. The city administration is interested in using WSNs for the early detection of brush fire events as well as to monitor their course. To accelerate the deployment of their new application and to avoid redundancy, the city administration has opted to deploy sensors in areas under its jurisdiction (i.e. streets and parks) and to re-use the WSN nodes already deployed in private homes. These sensors have several sensing capabilities, such as temperature, humidity, $CO_2$ and dust levels. They also execute multiple tasks (thanks to WSN virtualization), some of which may belong to semantic applications. The sensors, executing these tasks, provide annotated data for several semantic applications.

This sensor deployment can be utilized for several semantic applications. For example, the city administration's application can provide detailed information about fire events to its users, rather than simple notifications. Another example is of a weather applications that can use the same annotated data to identify prevailing weather condition such as sunny, haze, partial cloudy and snow. Similarly, a smart parking application could use the same annotated data to determine the current pollution levels and dynamically change the parking fee accordingly. For example, when the pollution level is very high, parking could be offered at a discount or even free.

### B. Requirements

Based on the scenario described above, we derive the following six requirements. *First*, the proposed architecture should allow for the real-time annotation of sensor data. This means that the sensor data should be annotated before sending it to the semantic applications. The *second* requirement is that the base ontology should be stored in the WSN in a distributed manner, since it will be used to annotate the sensor data. The *third* requirement is that the annotation should be done in a distributed manner without relying on a central node. This ensures that any node failure does not affect the annotation process. The *fourth* requirement is that it should be possible to enhance or to extend the ontology so that new concepts can be integrated with the existing ones. The *fifth* requirement is that the proposed solution should be applicable to heterogeneous sensor platforms and the data formats that they use, to ensure interoperability. The *sixth and final* requirement is the use of standardized ontologies, so that all semantic application can use standard concepts.

## III. PROPOSED ARCHITECTURE

In this section, we begin by discussing our previous architecture, since we use it as the basis for this work. Next, we present the architectural principles, followed by the details of layers and functional entities of the proposed architecture. Finally we present the base ontology that we used for sensor data annotation.

### A. Our Starting Point

The work in this paper is based on our previous WSN virtualization architecture [5] which is illustrated in Fig. 1. The architecture consists of four layers. The physical layer consists of sensor nodes that can run several application tasks simultaneously. Two types of sensor nodes are considered in the architecture. Type A sensors are resource-constrained sensors that have very limited processing and storage capabilities, e.g. TelosB motes. Type B sensors have better processing and storage capabilities, e.g. Java SunSpots. Since Type A sensors may not be capable enough to work together with other sensors in a group, they rely on more powerful nodes called Gate-to-Overlay (GTO) nodes for this purpose.

The virtual sensor layer abstracts the simultaneous tasks run by the physical sensors as virtual sensors. In this paper we use the terms virtual sensors and sensors interchangeably for consistency. To provide platform independence, the virtual sensor access layer consists of Sensor Agents (SAs). This independence is achieved by using standardized north-bound interfaces and proprietary south-bound interfaces. The final layer consists of application overlays that run simultaneously on top of the physical layer. There are separate interfaces for data and control messages. Overall, the architecture provides the flexibility of using multiple applications concurrently over WSN deployments.

### B. Architectural Principles

The *first* architectural principle is that the ontology used to annotate the sensor data is separated as base and domain ontologies. The former consists of concepts related to the deployed sensors and their capabilities, and is stored in the WSN, while the later consists of domain-specific, application-related concepts and is typically stored in the application domain. This basic principle allows the solution to become independent of any application domain.

The *second* architectural principle is that we use two independent overlays: one for data annotation and the other for storing the base ontology. Overlays have several advantages: they are distributed, they do not rely on centralized control and they allow resource sharing [10].

The *third* architectural principle is that every virtual sensor created for semantic application is represented in the annotation overlay by a corresponding entity that annotates its data.

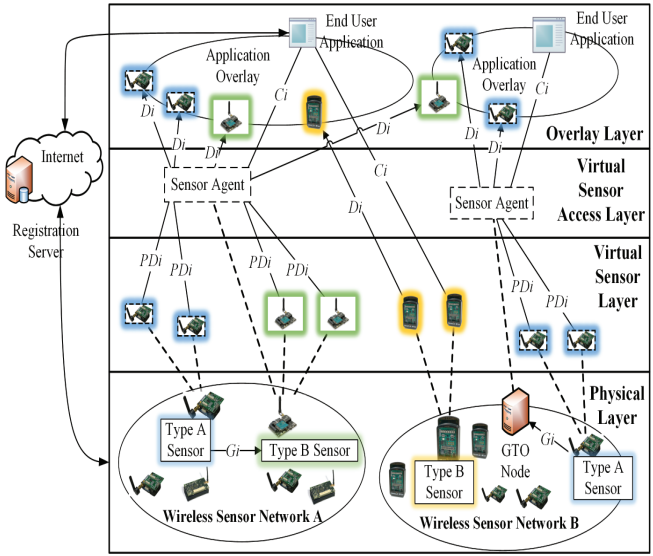

Fig. 1. WSN virtualization architecture

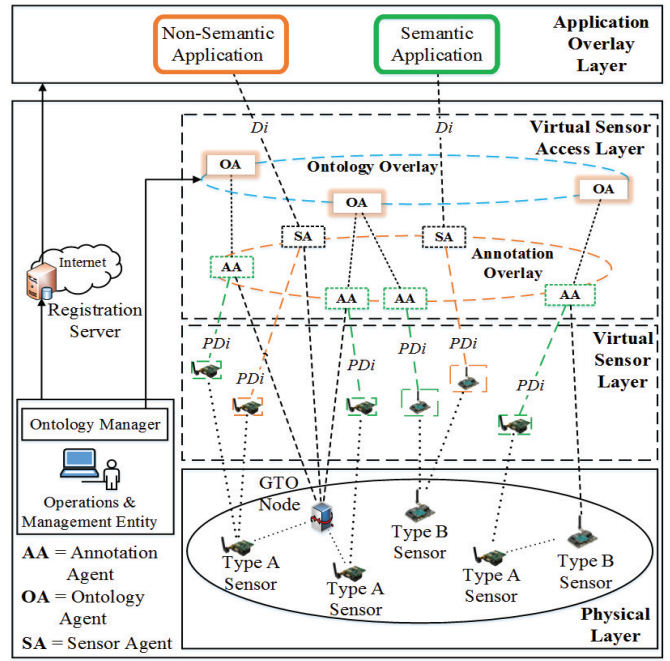

Fig. 2. Proposed data annotation architecture

This means that every sensor sending data to semantic applications will have a dedicated entity for annotation purposes.

The *fourth* principle is that, for resource constrained sensors, the annotations will be performed by capable sensors and other powerful nodes, e.g. gateways. This principle ensures that all kinds of sensors are available for the semantic applications.

*C. Layers and Functional Entities*

Fig. 2 shows the proposed architecture. It is based on our previous WSN virtualization architecture, presented in Section III-A. The physical layer remains the same while the virtual sensor layer now consists of two types of virtual sensors. The first group are the virtual sensors that are created for semantic applications, referred to hereafter as semantic virtual sensors. They are indicated as green-dashed boxes. The second type of virtual sensors are created for non-semantic applications, referred to hereafter as virtual sensors. These are shown as orange-dashed boxes. The difference between these two types of virtual sensors is that the raw sensor data from the green-dashed ones will be annotated before being sent to end-user semantic applications. The virtual sensor access layer has two new functional entities and two overlays. The functional entities are Annotation Agents (AAs) and Ontology Agents (OAs). We term an agent as an entity that provides a given functionality, therefore several agents are used in our architecture. The Annotation overlay consists of AAs, which annotate sensor data using the base ontology. They communicate with Sensor Agents (SA) in the same overlay to send the annotated data to the semantic applications. The Ontology overlay consists of OAs, which are responsible for storing the base ontology in a distributed manner. The OAs act as super-peers and provide the requested ontology to the AAs. They do not deal with the sensor data.

The architecture supports both semantic and non-semantic applications. The Operations & Management (O&M) entity, which is usually the infrastructure owner, is responsible for providing the base ontology. Since O&M entity is aware of the type of sensors deployed in the WSN, it can easily develop and disseminate the base ontology to the ontology overlay.

The architecture does not deal with the sensor discovery mechanism and storage of sensor data in a repository for data analysis. For the former, existing work such as [11], [12] can be reused. In this work we assume that the sensors have already been discovered and are stored in a registration server. For the latter, we leave it to the applications to decide on the sensor data storage since it is an application specific requirement.

The proposed architecture fulfills the set of requirements mentioned in Section II.B. AAs allow real-time annotation of sensor data in a distributed manner. OAs store the common ontology and are distributed using the concept of overlays. The base ontology can be extended by creating additional OAs. The architecture is platform-independent thanks to the SAs. As we use and extend SSN ontology in our work, the final requirement is also fulfilled.

*D. Base Ontology*

We have built our base ontology by extending the SSN ontology, since it is quite well-known and widely used to describe sensors and their data. As mentioned before, the goal of having a base ontology is to add metadata to the raw sensor data before it is used by a particular application. We assume that the WSN consists of temperature, humidity, light and carbon sensors and thereby incorporate these type of sensors and their observations in the base ontology. Fig. 3 shows the part of the base ontology, related to temperature sensors.

IV. PROCEDURES AND ILLUSTRATIVE SCENARIO

In our architecture we need different procedures related to the management and operational aspects of the annotation and ontology overlays. The management procedures include the following. 1) Selection of sensors and GTO nodes that will play the role of *i)* OAs in the ontology overlay, and *ii)* AAs in the annotation overlay. 2) The distribution of the ontology over the OAs. These procedures are motivated by our architectural principles mentioned in Section III.B.

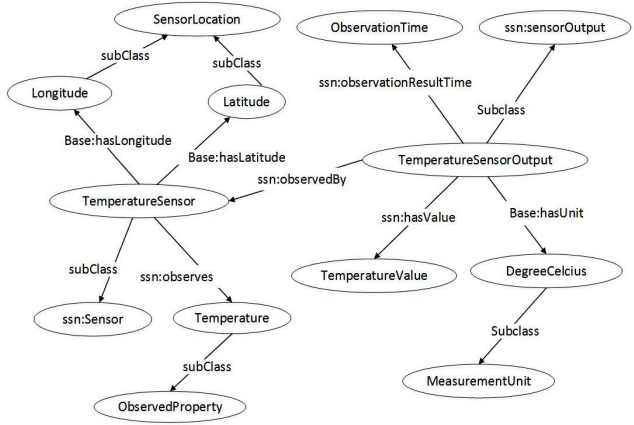

Fig. 3. Temperature sensor part of the base ontology

The annotation process requires the ontology, which may not be available with the AAs. This situation calls for an ontology discovery procedure to allow the AAs to annotate the sensor data. The operational procedures include the ontology discovery and the sensor data annotation.

### A. Management Procedures

According to the first and second architectural principles, we store the base ontology in the WSN using the concept of overlays, i.e. in the ontology overlay. The ontology overlay consists of OAs that require sufficient storage space and an efficient request/response mechanism. There are two types of nodes that can act as OAs: GTO nodes, which store the complete base ontology, and Type B sensors, which store part of the base ontology.

According to the third architectural principle, each sensor is represented by a corresponding AA in the annotation overlay. However, the role of an AA requires certain capabilities for computational-intensive tasks, such as the mapping sensor data to the base ontology concepts and generating output files. However, not all sensors are capable of performing these tasks, especially the ones that have 16-bit processors and memory on order of KBs, e.g., TelosB. For these sensors either Type B sensors or GTO nodes can act as AAs on their behalf, in accordance with the fourth architectural principle.

According to the second architectural principle, the base ontology needs to be distributed. The following mechanism is used for the distribution. GTO nodes contain the complete base ontology, while Type B sensors only contain the parts of the base ontology, related to phenomena that they sense. For example, a Type B sensor with temperature sensing capability will only contain the temperature portion of the base ontology. In order to accomplish this distribution, the GTO nodes split the base ontology into multiple parts and send it to the relevant Type B sensor. The common ontology concepts are present in each part. It is important to remember that since sensors are prone to failure, it makes sense to have the same parts of the base ontology present in multiple Type B sensors.

Both the GTO nodes and the Type B sensors can be selected for the roles of AAs and OAs. However, the OAs in the GTO nodes contain the complete base ontology, while the OAs in Type B sensors only contain the part of the ontology they require for annotation.

### B. Operational Procedures

The first operational procedure is the ontology discovery. There are two possible approaches, pro-active and reactive. In the pro-active approach OAs, as super-peers, periodically advertise the base ontology parts that they have. The AAs then send their ontology requests in response to these advertisements. In the reactive approach, AAs first determine the sensing capabilities of the corresponding sensors, based on which they send discovery request to their super-peers, for the required part of the base ontology.

The second operational procedure is the data annotation, which works as follows. The semantic virtual sensors send their data in a standardized or proprietary format to the AAs. Once an AA receives the raw sensor data, it first checks locally if it has the required ontology to annotate it, if not, a discovery request is sent to the ontology overlay. When it has the required ontology, the AA annotates the raw sensor data, and sends it to the SA. The SA is then responsible for sending the annotated data to the semantic application.

### C. Illustrative Scenario

The city administration and home owners deploy fire detecting sensors in public streets and in private homes, respectively. These sensors run multiple application tasks concurrently, using virtual sensors and semantic virtual sensors. The semantic virtual sensors send annotated data to the fire monitoring semantic application. The application receives this data and uses a reasoner to infer knowledge and to get detailed information about fire events.

The annotation process works as follows (a sequence diagram is presented in Fig. 4). Semantic virtual sensors send their raw data in a standardized or proprietary format to the AA. Once an AA receives the raw sensor data, it first checks locally to determine if it has the required ontology to annotate the data, if not it sends request message to an OA for the required ontology. The OA may request another OA for the required ontology if it does not store it. Once the ontology is retrieved, it is sent to the AA, which then annotates the raw sensor data using the received ontology and sends it to the SA. The SA sends the annotated data to the appropriate semantic application. The semantic application applies the domain ontology and a set of rules using a reasoner to infer additional knowledge. If a fire event is detected then a notification is sent to the end-user. The end-user may query for additional information such as fire status and location. In Fig. 4, the end-user queries for the fire status and receives the response, i.e. initial fire.

## V. PROTOTYPE IMPLEMENTATION AND RESULTS

In this section we present our prototype in detail. First we discuss the implementation choices we made, and then we present our prototype setup and the performance metrics. We end this section with a discussion of the results.

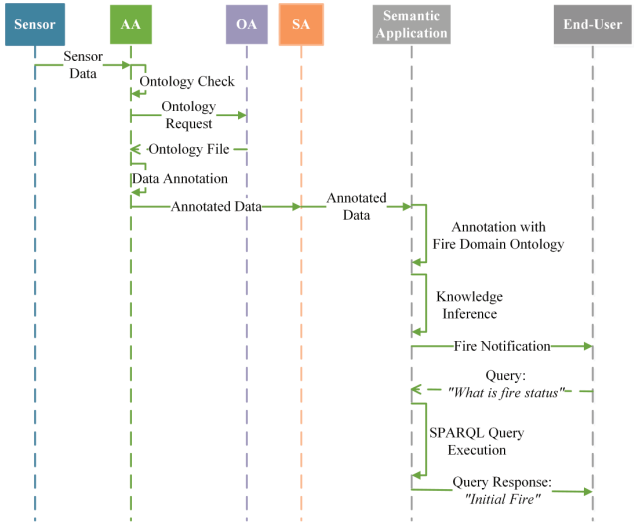

Fig. 4. Sequence diagram of the illustrative Scenario

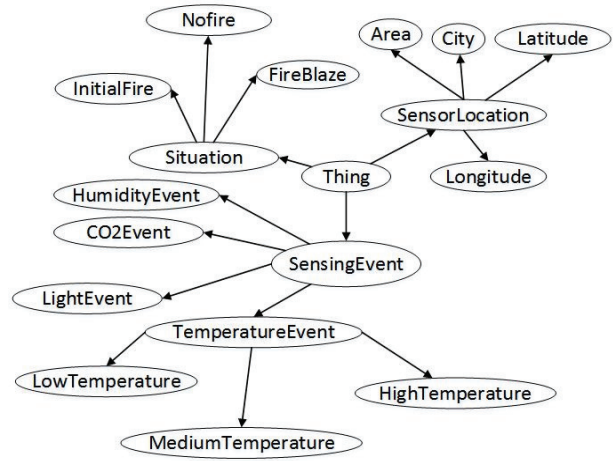

Fig. 5. Some concept of the fire domain ontology

### A. Implementation Choices

We developed a fire monitoring semantic application for our prototype based on the scenario presented in Section II.A. The application is offered as Software as a Service (SaaS) to the end-users. It was developed using the Apache Jena Framework, which is an open source Java framework for building semantic web and linked data applications. The application was deployed in a cloud-based Google App Engine (GAE), which is a Platform as a Service (PaaS) that allows the development of SaaS applications without having to maintain a server. We chose GAE because it makes it easy to deploy and maintain applications. The annotation and ontology overlays are implemented using the JXTA [13] protocol, an open source peer-to-peer protocol specification that allows the creation of independent, robust and efficient overlay networks.

The fire monitoring semantic application is a RESTful web service that uses the following components:

*1) Fire domain ontology:* Contains the concepts of fire, its states, and sensing events along with their states, such as temperature (high, low), relative humidity (high, low) levels, CO2 (high, low) levels and location (city, park, and downtown). Fig. 5 shows some concepts of the fire domain ontology.

*2) Jena Inference API:* Used to reason over the annotated data and to infer additional knowledge using a set of rules. We developed several rules for our semantic application to provide information to end-user about the fire events. Two examples of rules are given below.

```
[Rule1: (?output ssn:hasValue ?Value)
greaterThan(?Value,80), (?output rdf:type
base:TemperatureOutput),
(?output base:hasUnit base:DegreeCelsius) ->
(?output fda:hasTemperatureType:
fda:HighTemperature) ]

[Rule2: (?output fda:hasTemperatureType
fda:HighTemperature)
(?output fda:hasHumidityLevel fda:LowHumidity)
(?output fda:hasCO2Level fda:HighCO2)  ->
(?output fda:hasFireSituation fda: fireBlaze)]
```

*3) Query Engine:* Used to query annotated data. Below is an example query to get event information like event time, its value, location, and the status (fire event in this case).

```
SELECT  ?Time ?Temperature ?Longitude
?Latitude ?Firesituation
WHERE {
 ?SunSpotOutput base:hasSensingTime ?Time.
 ?SunSpotOutput ssn:hasValue ?Temperature.
 ?Sunspot base:hasLongitude ?Longitude.
 ?Sunspot base:hasLatitude ?Latitude.
 ?SunSpotOutput fda:hasFireSituation
?Firesituation.
 FILTER ( regex(str(?Firesituation),
'http://www.semanticweb.org/WirelessSensor/
 FireApplication#FireBlaze', 'i' )
 }
```

The functional entity AAs are in annotation overlay and have the following components:

*1) Web Server:* Receives the sensor data;

*2) JXTA Edge Peer:* Participates in the overlay and request the required parts of the base ontology;

*3) RDF Generator:* Annotates sensor data using the base ontology; and

*4) Web Client:* Sends annotated data to semantic application.

The functional entity OAs are in the ontology overlay and have the following component:

*1) JXTA Rendevous Peer:* To store the base ontology and send it to the requesting AA. We used the JXTA Content Management System (CMS) to advertise the base ontology available in each OA and send it to the requesting AAs.

The proposed architecture is implemented as Infrastructure as a Service (IaaS), which allows us to link our solution to the IaaS, PaaS and SaaS aspects of cloud computing paradigm.

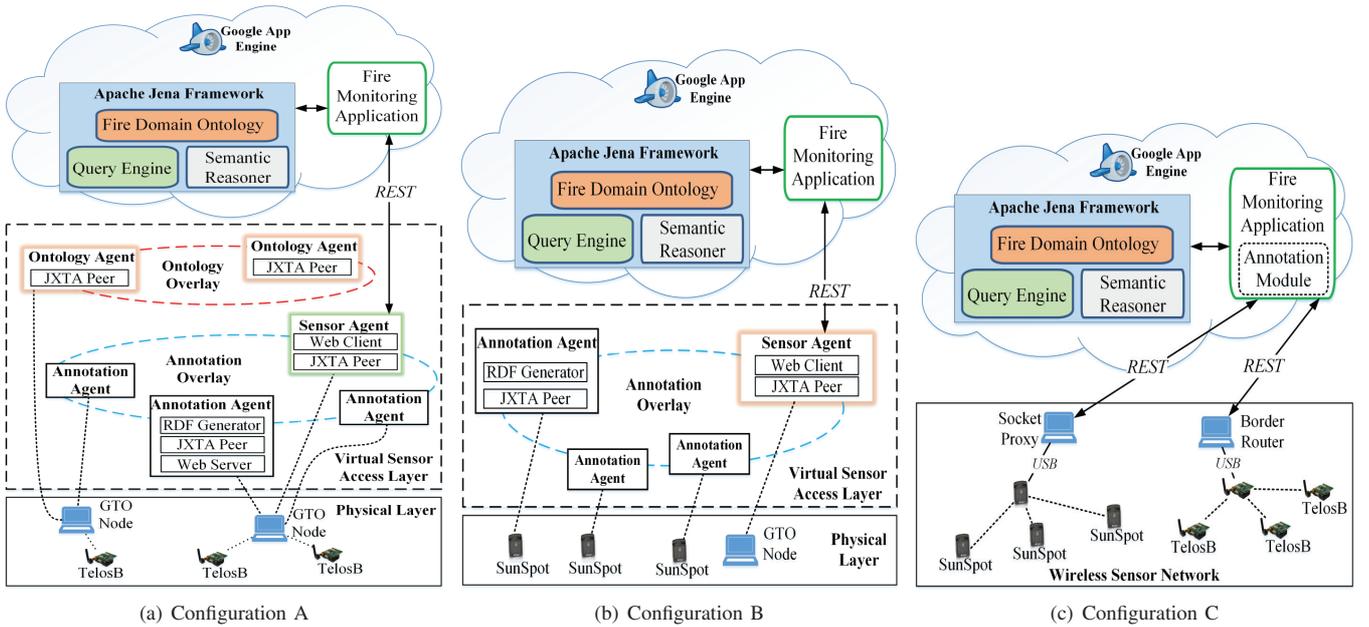

(a) Configuration A    (b) Configuration B    (c) Configuration C

Fig. 6. Implementation Architecture

## B. Prototype Setup

We used two different sensor kits for the prototype, Java SunSpot and TelosB motes from AdvanticSys Kit. In total we used 6 SunSpots (2 as base stations), 4 TelosB motes (1 as border router) running Contiki OS. All these sensors have multiple on-board sensing capabilities but differ in their processing and storage abilities. In our implementation, TelosB motes are Type A sensors and Java SunSpots are type B sensors. All of the sensor were running multiple application tasks. The Java SunSpots had three application tasks running concurrently, periodically measuring temperature, light and blinking LEDs. The TelosB motes had temperature, light, and humidity tasks running concurrently. Type B sensors send their data in SenML [14] format, which is a lightweight standard data model which is suitable for sending sensor data. Type A sensors send their data in simple string format. Fig. 6 shows the three implementation configurations we used for evaluation purposes. The details of these configurations are as follows:

*1) Configuration A:* We used Type A sensors (TelosB). The semantic virtual sensors sent their raw data to a GTO node. The GTO node (acting as an AA) downloaded the required ontology from an OA and annotated the raw sensor data. Lastly, the annotated data was sent to the fire monitoring semantic application via SA.

*2) Configuration B:* We used Type B sensors (Java SunSpots). The ontology used to annotate the data was stored locally in the Type B sensors, hence there is no ontology overlay. We implemented the AA in the Type B sensors using $\mu$Jena library [15]. This way they did not need any GTO node to perform annotation on their behalf. Each semantic virtual sensor generated the raw data, annotated it and sent it to the fire monitoring semantic application via SA.

*3) Configuration C:* We used both Type A and Type B sensors. All of the sensors sent their raw data over the Internet. For Type A sensors, we used a Contiki border router to allow them to directly communicate with the semantic application. For Type B sensors, we used Java Socket-Proxy which communicated with the semantic application on their behalf. In this configuration, the fire monitoring semantic application performed the annotation itself. This allowed us to measure the extra delay introduced by our approach.

## C. Performance Metrics

The prototype's performance was assessed in terms of the following metrics: End-to-End Delay (E2ED), Ontology Download Time (ODT), Impact of the scalability of AAs, Expected Operation Time (EOT) of Java SunSpots, and the Impact of tasks on current draw from Java SunSpots battery.

E2ED is the time difference between when the semantic virtual sensors sent their raw data and when the corresponding success code (200 OK) is received from the fire monitoring semantic application. It includes the time taken by all intermediate steps (i.e. receiving raw data at AA, ontology discovery and download (for configuration A), and annotation process). ODT is the time it takes an AA to request and to receive the required ontology from an OA. Impact of scalability of AAs was studied in terms of discovery of an OA and ODT. To find EOT of Java SunSpots, we executed both semantic and non-semantic tasks continuously until the Spots died. For this purpose no sleep or power saving mechanism was used. Finally we determined the current draw from Java SunSpot battery while in shallow-sleep mode (no task, radio ON), executing semantic, and non-semantic tasks. The experiments were repeated 50 times and their confidence interval is 95%.

## D. Results

Fig. 7 shows the individual E2ED of the three configurations. Configuration A has an average E2ED of *3566ms*. The actual annotation delay was negligible (*less than 10ms*), since the AA was implemented on a laptop computer. The E2ED of configuration B is the highest, at *4575ms*. The average annotation delay was *525ms*, since the Java SunSpots were annotating data themselves. We found that this longer time was attributable to the low RAM size, only 1MB. Despite this, SunSpots showed promise and were able to annotate sensor data and run other tasks concurrently without any other issues. The E2ED of configuration C is *3187ms*. As expected, the semantic application was able to annotate the sensor data quickly but at the expense of developing the base ontology and then implementing it in addition to the application logic. Fig. 8 shows their average E2ED of all configurations after 50 repetitions. The average ODT for configuration A is *94ms* as shown in Fig. 9, which is typical in LAN environment using JXTA protocol.

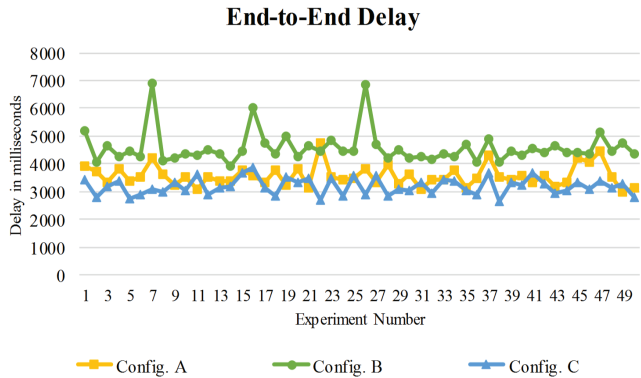

Fig. 7. End-to-End Delay

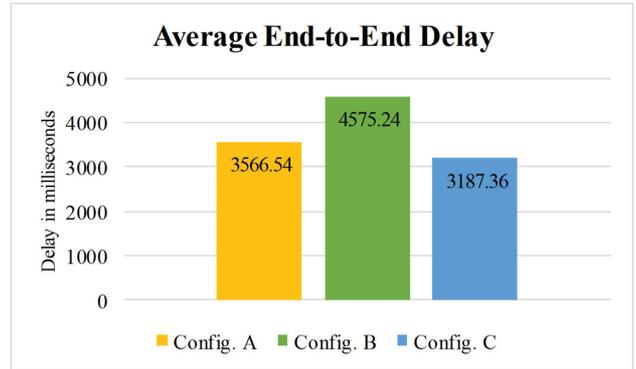

Fig. 8. Average End-to-End Delay

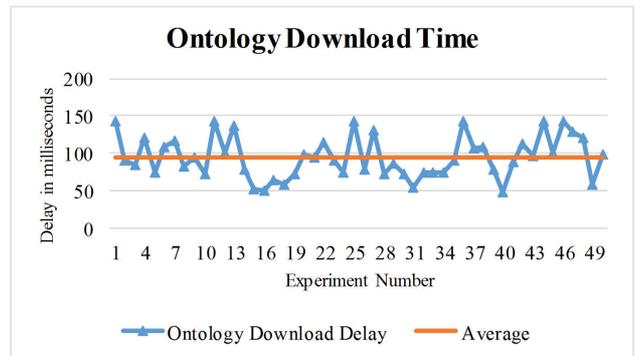

Fig. 9. Ontology Download Time

Since JXTA was used for implementation, it had direct impact on the scalability part. The results in Fig. 10 show the increase in OA discovery time when AAs increase. JXTA is known to perform poorly when peers in the network increase and this was demonstrated in this work. However, the increase in AAs did not impact the ODT mainly because OA was already discovered. Here the average ODT was around *100ms*, almost similar to the one shown in Fig. 9.

Fig. 11 shows the EOT of the Java SunSpots while running a semantic and a non-semantic task, without using any sleep mechanism. Without considering normal battery discharge, SunSpots last around *571* and *603* minutes operation time for the semantic and non-semantic tasks respectively. Using *0.8* as constant multiplier for normal battery discharge reduces the operation time to *456* and *482* minutes respectively. The SunSpots draw *38mA* current *(base value)* during the shallow mode (no task, radio ON), *75.6mA* for non-semantic task *(98% increase from base value)* and *79.8mA* for semantic task *(109% increase from base value)*.

For all three configurations, we also experienced delay due to circumstances beyond our control, e.g. from time to time GAE would start a new process for the fire monitoring semantic application and reload it thereby incurring unnecessary delay. We were able to determine this from the log files of our fire monitoring semantic application.

We believe that for future semantic applications, it will be important to use multiple WSN infrastructures that may not be geographically co-located. In such cases, it will be difficult to know beforehand, the capabilities of a WSN, the types of sensors and their observations. Also for WSN infrastructure owners may only want to share the sensor data instead of exposing their infrastructure altogether. In such situations, it makes sense to have an annotation mechanism that provides annotated data to multiple semantic applications.

## VI. RELATED WORK

A framework called semantic sensor web [7] annotates sensor data and provides situational awareness. The annotation is done using spatial, temporal and thematic metadata. In [16] the Sensor Observation Service SOS from SWE is extended by incorporating support for a semantic knowledge base. They use spatial, temporal and thematic ontologies to annotate sensor data. Both [7] and [16] rely on SWE, hence they are not suitable for resources-constrained environments. A two-layer architecture to annotate and query the sensor data is presented in [17]. The sensor data is collected in a pattern dictionary, in the back-end layer, to generate patterns along with semantic annotations. The patterns are used to determine the type of a new sensor and to automatically annotate its data. A crawler is used to retrieve the sensor data from multiple WSNs and store it after annotation. The front-end layer provides a GUI that the end-user utilizes to send search requests. The work is more focused on building automation domain.

In [18], the authors use their own SenMESO ontology for annotation which is a combination of various domain ontologies covering the sensor data and features of interest. The sensors send the observed data in SenML format to the gateways. The gateway nodes generate an XML file and send it to the aggregation gateways which use the stored ontologies to annotate the sensor data and thereby allow different applications to use it. As an extension of this work, the authors present a mechanism to annotate M2M data in [19]. The work focuses on developing semantic-based M2M applications. The authors designed an M3 ontology to integrate cross-domain M2M data. There are no details regarding network architecture, but a web-based prototype is available. Two cross-domain semantic-based applications are also discussed.

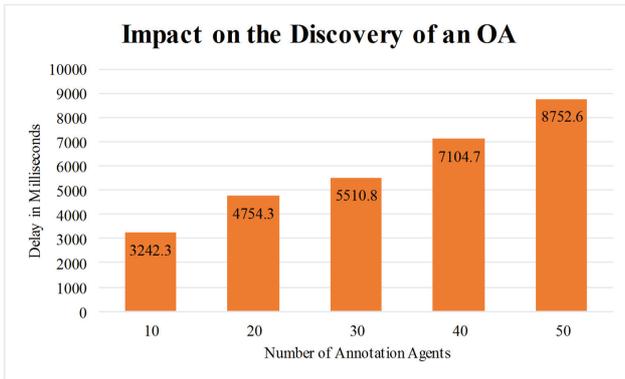

Fig. 10. OA Discovery Time When AAs Increase

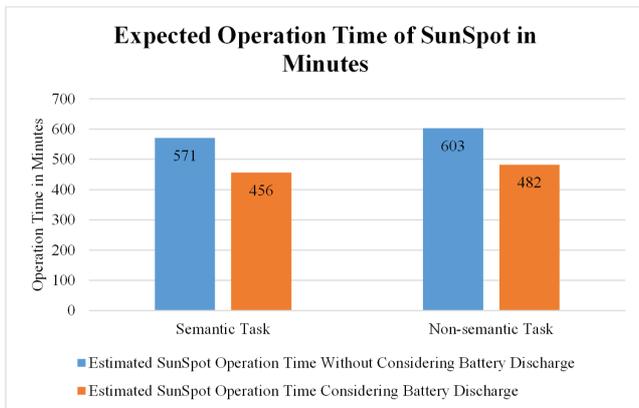

Fig. 11. Expected Operation Time of Java SunSpots (always on)

Overall, the existing studies have several limitations, such as domain-specific solutions, and their use of protocols such as Sensor Web Enablement (SWE) [9] that are difficult to setup and definitely not suitable for resource-constrained environments. Another limitation is that they are focused on interoperability between sensors rather than their data.

## VII. Lessons Learned and Future Work

We have learned several lessons. The first lesson is that WSN node-level virtualization is still a potent research area with very few solutions readily available. More efforts are required from designing a capable WSN operating system like [20] to unconventional energy harvesting mechanisms for sensor nodes like [21]. The second lesson is that current overlay middleware solutions are not suitable for WSNs because none has been designed to work with these resource-constraint devices. JXTA is too heavy for sensor nodes and its future is also uncertain. The third lesson is that there are not many libraries for semantic annotation that can be used by resource-constrained devices. We found an old J2ME-based $\mu$Jena library and after several modifications managed to use it with Java SunSpots. However it only annotates data in N-TRIPLE format, whereas standard Apache Jena Framework supports multiple formats. Extensions to $\mu$Jena library to annotate sensor data similar to Apache Jena Framework can be a useful contribution.

We have identified several key research issues that need to be addressed. First is the optimal selection of sensor nodes for the roles of AAs and OAs using energy-aware algorithms. These algorithms also need to take into account the characteristics of WSNs. Second issue is regarding the management of base ontology, since new types of sensors with new sensing capabilities may be deployed along with the existing WSN infrastructure. There is a need to have an easy to use mechanism to create and manage the ontology and later distribute it in the WSN infrastructure in an efficient manner. Third issue is that there is a need for lightweight P2P middleware for capable sensor nodes. This would make it possible for geographically-distributed sensors to share their data efficiently.

The final but very important issue is the possible integration of our proposed architecture with Platform-as-a-Service (PaaS) for the rapid provisioning of WSN application that can be offered as SaaS is yet another issue to investigate. In our current implementation we (partly) bypass Google Infrastructure for the interactions with our virtualized WSN infrastructure. As future work, we plan to integrate WSN infrastructure with a PaaS and allow its management at a higher level of abstraction through dynamic resource provisioning.

## VIII. Conclusion

Semantic applications are being used in many application areas such as life sciences, media, and information systems. Annotating sensor data allows the end-users to get high-level information about the real-world situations instead of raw measurements of individual sensors. This could potentially open doors to many new applications. In this paper we have proposed an architecture for annotating sensor data in virtualized WSNs where sensors run multiple application tasks concurrently. Our architecture is applicable to both resource-constrained and resource-full sensors. We have also demonstrated the feasibility of the proposed architecture by realizing a representative use case using heterogeneous sensors. Several research issues have also been identified as future work.

## Acknowledgment

This work is partially supported by CISCO systems through grant CG-576719, and by the Canadian Natural Science and Engineering Research Council (NSERC) through the Discovery Grant program.